# From Horseback Riding to Changing the World: UX Competence as a Journey

**Omar Sosa-Tzec**
Indiana University
901 E. 10th Street
Bloomington, IN 47408
omarsosa@indiana.edu

**Erik Stolterman Bergqvist**
Indiana University
901 E. 10th Street
Bloomington, IN 47408
estolter@indiana.edu

**Marty A. Siegel**
Indiana University
901 E. 10th Street
Bloomington, IN 47408
msiegel@indiana.edu

**ABSTRACT**
In this paper, we explore the notion of competence in UX based on a practitioner's perspective. As a result of this exploration, we observed four domains through which we conceptualize a *plane of sources of competence* that describes the ways a UX practitioner develop competence. Based on this plane, we present the idea of competence as a journey, whose furthest stage implies an urge towards transforming society and UX practice.

**ACM Classification Keywords**
H.5.m. Information interfaces and presentation (e.g., HCI): Miscellaneous.

## INTRODUCTION
There is no doubt that the notion of user experience (UX) is gaining momentum in the academic and professional domains. On one hand, during the last decade, many scholars in HCI have advocated for theories, frameworks, and methods in which the notion of *experience* is the main focus, embracing the complexity of the interrelation of *people,* the so-called *context of use,* and the computer systems in use [15]. Consequently, there has been a shift in HCI discourse towards culture, value, and participation [4]. On the other hand, many designers of software systems have adopted *user experience* (UX) as an umbrella term to describe different practices, methods, and approaches, through which they aim to provide pleasant and meaningful experiences of interaction with computer systems or interfaces [1,9,24]. The demand for competent UX designers has encouraged academic institutions and design firms to offer instruction in this "new" discipline. Moreover, it is possible to find a large amount of online and offline resources regarding UX, which focus on aspects of the process, product, or business. See the list in [3] as example. However, UX in this context seems to be a notion difficult to define, and it is still unclear what skills and competences a UX designer should have [2]. As a result, the desire to reconcile practice and research has increased [10,12,18,20,21,23]. Thus, some scholars have addressed the notion of UX competence from the pedagogical side, whereas others have advocated for an approach based on performance and experience in the workplace [11,13]. This paper makes a contribution to the latter approach by exploring what constitutes competence among UX practitioners and how they improve their competence [14].

We identify four *domains of competence* from interviewing UX practitioners. Positioned in these domains are *sources of competence,* activities and fields from which a competence is gained. The domains describe the role of such a source: whether it belongs to a designerly discipline or not, and whether it focuses on aspects of implementation or management in UX practice. Then, we present a formulation of *competence as a journey* derived from the analysis of these practitioners' responses, which help us glimpse why not only technological or technical sources of competence should be accounted for when discussing and researching UX competence and what the value may be for a practitioner to embrace sources of competence foreign to the discourse that defines the discipline.

This paper is structured as follows. First, we describe our research approach. Then, we present our findings, expressed in terms of what we call *domains of competence*. We end the paper with a discussion based on such domains.

## RESEARCH APPROACH
In order to gain a better comprehension of competence in UX, we carried out an exploratory study. We are aware that competence and expertise are two notions that have been widely studied [5,7,11,17,19]. However, this study was not intended to bring a new formulation to the notion, or to test a specific proposition of any previous formulation of competence, or expertise for that matter. Instead, we started with a notion of competence derived from everyday practice as an attempt to see what possible aspects have been overlooked in other scholars' work on UX competence.

This motivated us to conduct an exploratory study through which we could gain insights and develop new questions concerning UX and design competence. Our key consideration for this study was to compare and contrast perspectives from UX-centered practitioners working inside and outside of the United States. We interviewed 8 practitioners, 3 female and 5 male. The interviewees were chosen based on certain criteria related to professional competence. We wanted professionals that had attained a certain position as a competent practitioner and that had significant professional experience. 4 of 8 practitioners work in the United States, and most of them are associated with a large American software company. The other 4 are located in the European and American continents, and they have experience with UX projects of international scope. Besides Spanish, their native language, they are proficient in English, and 3 also speak Portuguese. On average, the 8 practitioners had 11 years of experience. 1 of them had only 2.5 years of experience at the moment of the interview, but the rest ranged from 7 to 18 years. The 4 practitioners working in the United States reported holding a graduate degree, namely, HCI, Human Factors, Communication and Digital Media, and Design. The remaining 4 practitioners reported holding an undergraduate degree only, namely, Information Design, and a dual degree in Political Sciences and Sociology.

Two researchers conducted semi-structured interviews, one for each set of practitioners. For the practitioners working in the United States, the interviews were conducted *in situ*, whereas the remaining 4 were conducted online. In all cases, audio was recorded and notes taken. The duration of the interviews ranged from 1 to 2 hours. We asked the practitioners to explain their notion of competence. Additional questions focused on the themes: *assessing competence, improving competence, sharing competence, academic influence,* and *extended competence.* After the interviews, one of the researchers performed a thematic analysis on the notes and transcripts from the interviews of the practitioners working outside the United States. After discussion with the other author to reconcile the themes of this set of responses, this first researcher performed a thematic analysis on the remaining set of responses, seeking in part, a correspondence with themes found in the first set.

**DOMAINS OF COMPETENCE**
We found similarities in both groups regarding the notion of competence, including the experiences from which such a notion is built. Basically, all the interviewed practitioners described *competence* as the power to accomplish what is expected for them to do. Competence entails performing the entrusted job. Regardless of the years of experience, these practitioners recognize their own *degree of competence* and they affirm feeling good about it. Nevertheless, they have also experienced a lack of competence on certain occasions. For them, competence becomes visible when they demonstrate that they have the experience, knowledge, methodology, and skills for overcoming a situation within a project. This includes establishing connections and finding appropriate forms for the communication and deployment of information for stakeholders in that project, whether they are related to design or not.

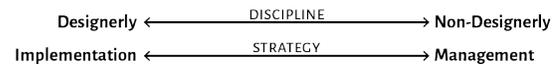

**Figure 1. Continua of sources of competence.
Figure by the authors.**

The initial analysis of the interviews led us to shift our primary focus to the notion of a *source of competence*. In the responses, we noticed that knowledge, inspiration, insights, and skills do not only come from design-focused learning resources. These practitioners include other sources of competence that are related to their everyday life and closer to their personal character. Through our analysis, we identified two continua describing sources of competence. The *discipline* continuum ranges from the *designerly* to the *non-designerly* sources of competence, whereas the *strategy* continuum ranges from sources of competence related to *implementation* to those related to *management* (Fig. 1). Each of the extremes of these continua defines what we call a *domain of competence,* which corresponds to one of the four emergent categories derived from our analysis. Each of them is explained below.

**Designerly domain**
This domain encompasses sources of competence generated in the diverse design-centered disciplines, such as graphic design, architecture, and industrial design. From these sources of competence, UX practitioners select and appropriate the theories, methods, and tools through which they not only develop *skills for designing,* but also help them to shape their *identity as designers. Adam,* one of our interviewees, considers that knowing *formal principles* is relevant for his praxis and to teach design to others. He also considers *history of design* as a factor for being competent, because learning about changes in ideology and modes of production during the last 100 years could help in reflections upon the scope and lifespan of an app, for example.

This domain also encompasses sources of competence required for design-oriented practices focused on interactive systems and digital-laden services that fall under the umbrella of UX. Specialized books, blogs, magazines, conferences, and even meet ups are included in this domain. As we noted, in early stages of competence development, practitioners seem to focus on the people and sources of competence that could help them acquire *best UX practices*. With the foundations established, new criteria might appear, whose purpose is fit to the practitioner's character or current professional context. This includes *curating* content from other design disciplines or paying attention to the *local UX discourses.* For example, *Adam* and *David* commented that J.J. Garret's "Elements of the User Experience" was a foundational resource in the early stages of their careers. Now senior designers, they have established new criteria in order to stay competent. For instance, Adam reads design books whose content has *stood the test of time.* In turn, David looks for LinkedIn and Facebook groups in order to learn about *the trend* established by the local UX community of the city wherein he currently works. For him, this is an organic activity to learn about people, companies, and *terminology* employed by such a community. In general, most of the interviewees see networking as a way to stay competent.

**Non-designerly domain**
This domain encompasses sources of competence generated outside of what could be considered as a design discipline, such as anthropology, psychology, business, marketing, and computer science. From these disciplines, UX practitioners select and appropriate frameworks, methods, and tools, and thus develop skills through which they demonstrate competence as well. For example, *Cameron* commented on "spending half of the time programming and half of the time learning about it" because it helps to increase his personal awareness of the limitations that both computers and programming have, which has influenced his perspective of UX design. However, Cameron argues that his design background has allowed him *to break barriers* that programmers would not dare to break, and that investing in programming over 10 years will help him comprehend the foundations of programming. He expects that after these years, paying attention to programming will be unnecessary, and he will be interested in *mastering* something else.

In this domain, there are also sources of competence applied to professional practice that comes from experiences and activities occurring *outside the workplace. Adam* commented on paying attention to *peripheral disciplines* as a way to improve UX practice. As he explained, reading about ant colonies or ecosystems could help a designer understand the behavior and growth of a group of users. *David* explained that taking classes on horseback riding and practicing the DeRose Method has influenced in his performance as a project leader. Horseback riding has *taught* him how to *lead* and *direct* people, whereas the DeRose Method has made it possible for him to endure the long working hours. *Emily* commented on having attended conferences about drug therapies and educational measurement in order to find people capable of *seeing the world differently* and to know whom to bring in during the planning of innovation projects. Other sources of competence mentioned by the practitioners include parenting, traveling, playing musical instruments, and cooking. Although these topics might be seen as mundane or unrelated to their professional practice, all the interviewees recognized the relevant role of such experiences and activities to stay competent in UX.

**Implementation domain**
The implementation domain encompasses sources of competence embodied in frameworks, methods, tools, and demonstrable in skills that UX practitioners utilize for making a design *concrete,* whether such competence could be regarded as designerly or not. *Information literacy* is a relevant competence for most of the interviewees. Knowing *what* to look for and *where* to find appropriate and applicable knowledge for the current situation is important to show competence. *Emily* remarked, "I can't recite [the usability heuristics], but I know when to bring them in," which exemplifies this competence. As we noted, *challenges* and *constraints* from each project shape the development of competence. Moreover, experiencing *failure* is a possibility, which helps practitioners become aware of the *scope* of their competence. For example, *Barry* and his colleagues decided to learn *Ruby* and program by themselves an intranet, for which they were commissioned, since they had access to learning resources about this programming language. However, they showed a lack of competence during the presentation to the client, since their prototype had plenty of *bugs. Cameron* reported having lost the database of one of his personal projects, because he was in charge of his own database server and code. Later, *appropriation* and *adaptation* might be competences that complement being information literate. For example, *David* pointed out that practice gives no time for learning heavy theories. What David seeks in resources is a framework that might work for the current situation and then adapts it. *Helen,* who made a similar case, added that *tweaking* a method or process is also necessary to fit in with other people's non-design backgrounds, since the main goal is getting things done.

**Management domain**
This domain encompasses sources of competence involved with *decisions* taken by UX practitioners, which aim at professional, business, or social impact in the short, medium, or long term. As it happens in the implementation domain, the sources of competence in the management domain could be designerly or not. For example, *Gloria,* with only a couple of years of experience, has become aware of the relevance of creating an impact at a business level. She is figuring out "which information can be delivered to which people and [in] which way," with the purpose of fitting in with the *business value* and getting promoted in the company for which she works. Interviewees with more than 10 years of experience made a similar case. *Emily* relies on her competence for managing people and asking the right questions. Additionally, she *looks for novelty* in order to cause a business impact and get funded, which she achieves by establishing multidisciplinary partnerships. According to *Adam,* the more senior you become, the less afraid you are of asking the appropriate question to the top-level or most knowledgeable stakeholder in the project. *Communication* and *Awareness of the place of design in relation to other areas* in a project are two important competences for Adam, which he instructs to his UX design staff. For most of the interviewees, *mentoring* is a competence related not only with learning, instructing, or even advocacy, but also with establishing *leadership,* creating *empathy,* and gaining *recognition.* As we observed, *networking* can play these roles in some occasions. With more experience, UX practi-tioners may want to achieve a social impact. For example, *David* thinks that a major human and social concern is *coming,* not only for business, but also for products, services and strategies. Based on his belief that traditional business models will not work for *social innovation, David* wonders how to transform what he has learned from working for different companies into a *sustainable social consulting firm* that really pursues people's *well-being* without charity-based funding. Further, David's vision involves utilizing his knowledge in interaction design to create a *tool kit for non-designers* interested in *creating companies* and materializing *solutions* for social innovation.

*mature design thinking,* a practitioner might start considering the creation of new business opportunities for causing a social impact or influencing the discipline. This would be the *ultimate competence,* a stage that we interpret as a practitioner's aspiration to *break paradigms* within her or his near future context.

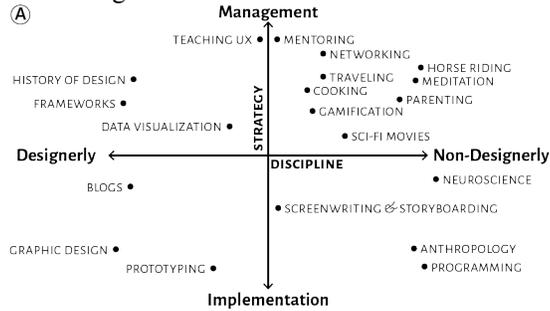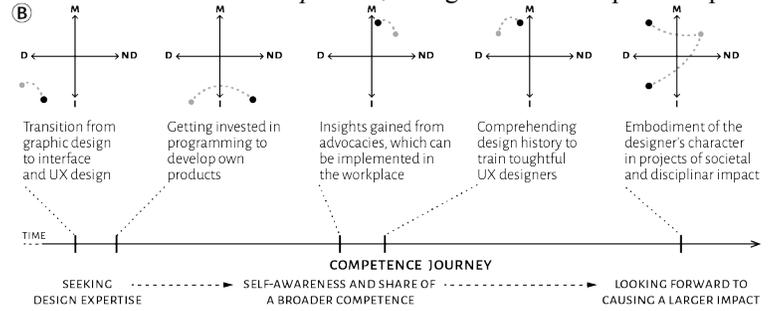

**Figure 2. (A) Plane of sources of competence. (B) Representation of competence as a journey. Figure by the authors.**

## DISCUSSION

Through the analysis of the responses from the practitioners that participated in this study, we noticed that the four emergent categories from above are not exclusive. Together, they form a plane with four quadrants, which help characterize an overall source of competence in a metaphorical sense. The four quadrants that compose this *plane of sources of competence* are *non-designerly management, non-designerly implementation, designerly implementation,* and *designerly management.* We have placed some of sources of competence we identified in the plane (Fig. 2A). This approach makes it possible to get an overview of the interviews and to get a sense of what quadrants were more mentioned than others. For instance, it may be interesting and even surprising that the *non-designerly management* quadrant has the most sources mentioned. Another possibility that this plane opens up to is to explore and plot certain *paths* through it. We can for instance ask the question if it is possible to find *a common path* in most of the cases of UX practitioners when it comes to how they have developed their competence.

Based on our interviews and the stories we heard from our interviewees, it is clear that they experience their competence development over time. We see these paths as *moments* in a *competence journey* (Fig. 2B). The typical journey of a UX practitioner would begin with the acquisition of technical and foundational competences for considering oneself as a designer and a UX practitioner. In later stages, the practitioner would start exploring sources of competence outside of the realm of design, and thus extrapolate her or his competence into a broader and more adaptable perspective of design. Nevertheless, and maybe related to

Along this *competence journey*, the plane of sources of competence would show *traces* that a practitioner leaves while moving from one source of competence to another. The practitioner's *curiosity* and *awareness*, and other *intrinsic* and *extrinsic forces* would motivate this movement. Further, traces going from the *non-designerly* towards the *designerly* half of the plane might talk about the link between the practitioner's professional and personal character, and about the passions, beliefs, and values that she or he might not only embody in objects, but also inculcate in others, whether they are UX designers or not.

The competence journey is similar to other interpretations of competence development as an *evolving* or *ongoing* process, a transition from being a *novice* to becoming a *master,* or even more, a *visionary* [7,13,17,19,22]. This journey, which is not formulated to focus on the transformation of mental functions [7], emphasizes a practitioner's experiential knowledge and its origin, whether it is gained from the professional practice or work outside the workplace. This interpretation resonates with approaches to UX competence focused on the development and share of competence at a personal level [11]. However, it also considers the goal of becoming a *visionary* [17], a person moving across domains and seeking a large impact, as a key characteristic for UX to be considered a designerly practice [5,6,8,16,25]. Furthermore, it emphasizes the non-linear linkage between the practitioner's personal and professional characters.

The competence journey, centered on the role of different sources of competence over time, suggests a greater fusion between the *personal* and the *professional* the more *senior* a practitioner becomes. It suggests an urge for *designing* oneself and the context when the *ultimate competence* has been reached. To become *true* designers, practitioners need to leverage non-designerly advocacies and experiences, an aspect that UX research and pedagogy should consider.

**ACKNOWLEDGMENTS**

This work is supported in part by the National Science Foundation (NSF) Grant Award no. 1115532. Opinions expressed are those of the authors and do not necessarily reflect the views of the entire research team or the NSF.

**REFERENCES**

1. Gavin Allanwood and Peter Beare. 2014. *User experience design: creating designs users really love*. Bloomsbury Academic, London ; New York.
2. Jonathan S. Arnowitz. 2015. The User Experience Designer's Charlatan Test: A First Step Towards UX Sanity Checking. *Proceedings of the 33rd Annual ACM Conference Extended Abstracts on Human Factors in Computing Systems*, ACM, 517–529. http://doi.org/10.1145/2702613.2732503
3. Chris Bank. UX Design Blog and Resources to Follow Religiously. Retrieved September 22, 2015 from http://www.awwwards.com/ux-design-blog-and-resources-to-follow-religiously.html
4. Susanne Bødker. 2015. Third-wave HCI, 10 Years Later—participation and Sharing. *interactions* 22, 5: 24–31. http://doi.org/10.1145/2804405
5. Nigel Cross. 1982. Designerly ways of knowing. *Design studies* 3, 4: 221–227.
6. Kees Dorst. 2006. *Understanding design: [175 reflections on being a desinger]*. BIS, Amsterdam.
7. Stuart E. Dreyfus and Hubert L. Dreyfus. 1980. *A Five-Stage Model of the Mental Activities Involved in Directed Skill Acquisition*.
8. Daniel Fallman. 2003. Design-oriented Human-computer Interaction. *Proceedings of the SIGCHI Conference on Human Factors in Computing Systems*, ACM, 225–232. http://doi.org/10.1145/642611.642652
9. Jesse James Garrett. 2011. *The elements of user experience: user-centered design for the Web and beyond*. New Riders, Berkeley, CA.
10. Elizabeth Goodman, Erik Stolterman, and Ron Wakkary. 2011. Understanding Interaction Design Practices. *Proceedings of the SIGCHI Conference on Human Factors in Computing Systems*, ACM, 1061–1070. http://doi.org/10.1145/1978942.1979100
11. Colin M. Gray. 2014. Evolution of Design Competence in UX Practice. *Proceedings of the SIGCHI Conference on Human Factors in Computing Systems*, ACM, 1645–1654. http://doi.org/10.1145/2556288.2557264
12. Colin M. Gray, Erik Stolterman, and Martin A. Siegel. 2014. Reprioritizing the Relationship Between HCI Research and Practice: Bubble-up and Trickle-down Effects. *Proceedings of the 2014 Conference on Designing Interactive Systems*, ACM, 725–734. http://doi.org/10.1145/2598510.2598595
13. Colin M. Gray, Austin L. Toombs, and Shad Gross. 2015. Flow of Competence in UX Design Practice. *Proceedings of the 33rd Annual ACM Conference on Human Factors in Computing Systems*, ACM, 3285–3294. http://doi.org/10.1145/2702123.2702579
14. Paul Hager and Andrew Gonczi. 1996. What is competence? *Medical Teacher* 18, 1: 15.
15. Steve Harrison, Deborah Tatar, and Phoebe Sengers. 2007. The three paradigms of HCI. *Alt. Chi. Session at the SIGCHI Conference on Human Factors in Computing Systems San Jose, California, USA*, 1–18. Retrieved July 15, 2015 from http://people.cs.vt.edu/~srh/Downloads/HCIJournalTheThreeParadigmsofHCI.pdf
16. Jon Kolko. 2015. *Exposing the magic of design: a practitioner's guide to the methods and theory of synthesis*. Oxford University Press, New York.
17. Bryan Lawson and Kees Dorst. 2009. *Design expertise*. Elsevier, Architectural Press, Oxford.
18. Jonas Löwgren and Erik Stolterman. 2004. *Thoughtful interaction design a design perspective on information technology*. MIT Press, Cambridge, Mass. Retrieved September 24, 2015 from http://site.ebrary.com/id/10405262
19. Harold G. Nelson and Erik Stolterman. 2012. *The design way: intentional change in an unpredictable world*. The MIT Press, Cambridge, Massachusestts ; London, England.
20. David J. Roedl and Erik Stolterman. 2013. Design Research at CHI and Its Applicability to Design Practice. *Proceedings of the SIGCHI Conference on Human Factors in Computing Systems*, ACM, 1951–1954. http://doi.org/10.1145/2470654.2466257
21. Yvonne Rogers. 2004. New theoretical approaches for human-computer interaction. *Annual Review of Information Science and Technology* 38, 1: 87–143. http://doi.org/10.1002/aris.1440380103
22. Martin A. Siegel and Erik Stolterman. 2009. Metamorphosis: Transforming Non-designers into Designers. Retrieved September 22, 2015 from http://shura.shu.ac.uk/449/
23. Erik Stolterman. 2008. The nature of design practice and implications for interaction design research. *International Journal of Design* 2, 1: 55–65.
24. Russ Unger (ed.). *A Project Guide to UX Design*.
25. Tracee Vetting Wolf, Jennifer A. Rode, Jeremy Sussman, and Wendy A. Kellogg. 2006. Dispelling "Design" As the Black Art of CHI. *Proceedings of the SIGCHI Conference on Human Factors in Computing Systems*, ACM, 521–530. http://doi.org/10.1145/1124772.1124853